\PassOptionsToPackage{prologue,dvipsnames}{xcolor}
\documentclass[sigconf]{acmart}
\AtBeginDocument{%
  \providecommand\BibTeX{{%
    \normalfont B\kern-0.5em{\scshape i\kern-0.25em b}\kern-0.8em\TeX}}}

\usepackage[capitalise]{cleveref}
\usepackage{hyperref}
\usepackage[dvipsnames]{xcolor}
\usepackage{fancyvrb}
\usepackage{circledsteps}
\pgfkeys{/csteps/inner color=white}
\pgfkeys{/csteps/outer color=white}
\pgfkeys{/csteps/fill color=MidnightBlue}

\begin{document}

\title{NLQxform-UI: A Natural Language Interface for Querying DBLP Interactively}

\author{Ruijie Wang}
\orcid{0000-0002-0581-6709}
\affiliation{%
  \institution{Department of Informatics\\ University Research Priority Program\\ ``Dynamics of Healthy Aging''}
  \institution{University of Zurich}
  \streetaddress{Zurich}
  \country{Switzerland}
}
\email{ruijie@ifi.uzh.ch}

\author{Zhiruo Zhang}
\orcid{0009-0008-7115-9429}
\affiliation{%
  \institution{Department of Informatics}
  \institution{University of Zurich}
  \streetaddress{Zurich}
  \country{Switzerland}
}
\email{zhiruo.zhang@uzh.ch}

\author{Luca Rossetto}
\orcid{0000-0002-5389-9465}
\affiliation{%
  \institution{Department of Informatics}
  \institution{University of Zurich}
  \streetaddress{Zurich}
  \country{Switzerland}
}
\email{rossetto@ifi.uzh.ch}

\author{Florian Ruosch}
\orcid{0000-0002-0257-3318}
\affiliation{%
  \institution{Department of Informatics}
  \institution{University of Zurich}
  \streetaddress{Zurich}
  \country{Switzerland}
}
\email{ruosch@ifi.uzh.ch}

\author{Abraham Bernstein}
\orcid{0000-0002-0128-4602}
\affiliation{%
  \institution{Department of Informatics}
  \institution{University of Zurich}
  \streetaddress{Zurich}
  \country{Switzerland}
}
\email{bernstein@ifi.uzh.ch}

\renewcommand{\shortauthors}{Wang et al.}

\begin{abstract}
In recent years, the DBLP computer science bibliography has been prominently used for searching scholarly information, such as publications, scholars, and venues.
However, its current search service lacks the capability to handle complex queries, which limits the usability of DBLP.
In this paper, we present NLQxform-UI, a web-based natural language interface that enables users to query DBLP directly with complex natural language questions.
NLQxform-UI automatically translates given questions into SPARQL queries and executes the queries over the DBLP knowledge graph to retrieve answers.
The querying process is presented to users in an interactive manner, which improves the transparency of the system and helps examine the returned answers.
Also, intermediate results in the querying process can be previewed and manually altered to improve the accuracy of the system.
NLQxform-UI has been completely open-sourced: \href{https://github.com/ruijie-wang-uzh/NLQxform-UI}{\color{blue} https://github.com/ruijie-wang-uzh/NLQxform-UI}.

\end{abstract}

\ccsdesc[500]{Information systems~Question answering}
\ccsdesc[500]{Information systems~Search interfaces}
\ccsdesc[500]{Human-centered computing~Natural language interfaces}

\keywords{Knowledge Graph Question Answering, Natural Language Interface, SPARQL, Language Model}

\maketitle

\section{Introduction}
\label{section_introduction}

The DBLP computer science bibliography\footnote{\url{https://dblp.org/}} is one of the most widely used scholarly databases and web services, covering over 4.4 million publications and more than 2.2 million scholars.\footnote{https://dblp.org/faq/1474565.html}
The rich metadata of DBLP could support answering complex queries that involve multiple resources (e.g., multiple publications and scholars), relationship constraints between resources (e.g., authorship and affiliation), and operations over query results (e.g., count and time-based filtering).
Example queries include ``\textit{enumerate the authors of Attention is All You Need along with other papers they published},'' and ``\textit{what papers has Tim Berners-Lee published in the last 5 years?}''
These complex queries could be frequently asked by researchers in daily research or literature reviews.
However, the current search service of DBLP\footnote{\url{https://dblp.org/search/}} is based on string matching and cannot handle them.
Researchers would have to manually search the topic resources of the questions (e.g., \textit{Attention is All You Need}), integrate the search results, and potentially repeat this process multiple times to find the answers. Note that other prominent literature search services, such as Google Scholar\footnote{\url{https://scholar.google.com/}} and ACM Digital Library,\footnote{\url{https://dl.acm.org/}} also cannot directly answer this kind of complex queries that involve multiple joins and/or aggregations.

In this paper, we present \textbf{NLQxform-UI}, a web-based natural language \textbf{U}ser \textbf{I}nterface powered by our previous question answering (QA) system \textbf{NLQxform}~\cite{DBLP:conf/semweb/0003ZRRB23}.
NLQxform is named after its \textbf{transform}er-based \textbf{transform}ation of \textbf{N}atural \textbf{L}anguage \textbf{Q}uestions.
Specifically, it leverages BART~\cite{DBLP:conf/acl/LewisLGGMLSZ20} to transform questions into executable SPARQL\footnote{\url{https://www.w3.org/TR/sparql11-query/}} queries.
The transformation involves four steps: 
\textbf{Step-I BART-based question to logical form translation}, which translates a given question into a logical form~\cite{DBLP:conf/semweb/0003ZRRB23} that has the same structure as the target SPARQL query but is still partially based on natural language expressions, 
\textbf{Step-II DBLP Search-based entity linking}, which links the expressions of entities in the logical form to entity URLs in DBLP,
\textbf{Step-III SPARQL template-based query correction}, which utilizes a pre-constructed SPARQL template base~\cite{DBLP:conf/semweb/0003ZRRB23} to correct potential minor errors in the logical form,
and \textbf{Step-IV: SPARQL endpoint-based answer retrieval}, which generates final target queries and executes them over the DBLP knowledge graph (DBLP KG)\footnote{\url{https://blog.dblp.org/2022/03/02/dblp-in-rdf/}} to retrieve final answers.

\begin{figure}[t]
    \centering
    \includegraphics[width=0.9\linewidth]{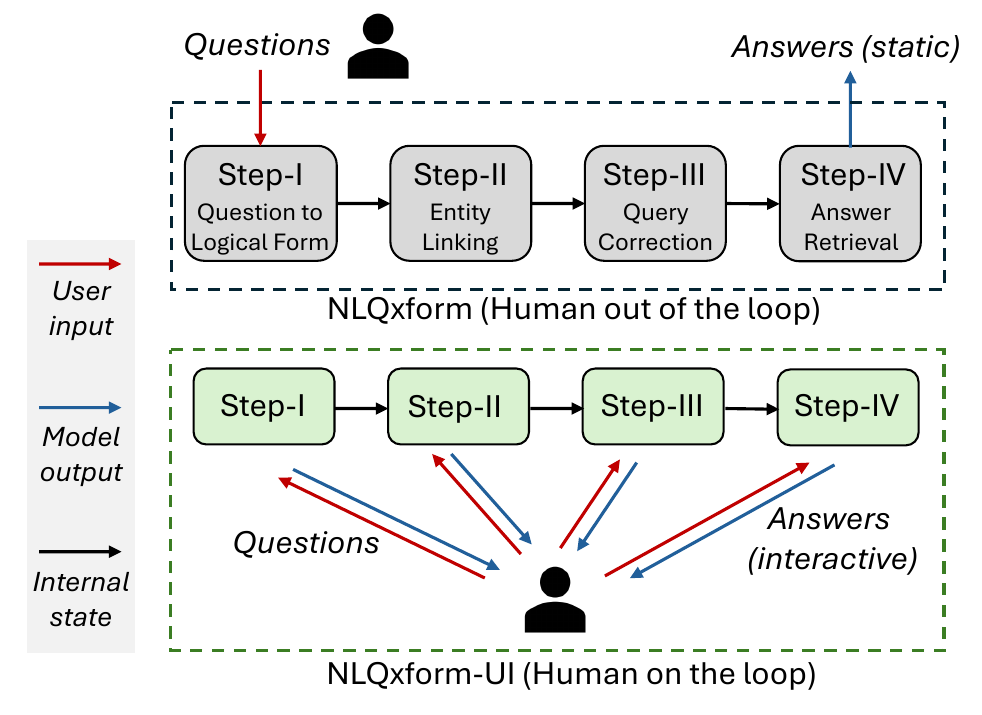}
    \caption{A diagram highlighting the difference between the existing NLQxform model and the proposed NLQxform-UI system.} 
    \label{figure_difference}
\end{figure}

NLQxform-UI leverages NLQxform to answer questions in a similar manner via the above four steps.
However, as a web-based interactive system, NLQxform-UI presents several new contributions and strengths.
We highlight the difference between NLQxform and NLQxform-UI in \cref{figure_difference}, where the red arrowed line signifies the input from the user (e.g., manual editing or selection of provided options), the blue arrowed line indicates the model output that is visible to the user in an intuitive way (e.g., final answers or previews of intermediate results), and the dark arrowed line represents the internal state of the model that is invisible or hard to read for the user (e.g., various data structures).
The original NLQxform model is a black box for the user, as the querying process is an invisible linear pipeline.
Therefore, examining the querying process and the returned answers is difficult.
Also, the answers of NLQxform are static texts, such as a list of URLs that the user needs to manually open to check.
In contrast, the querying process of NLQxform-UI is a human-on-the-loop process, where the intermediate results are presented to the user intuitively and can be interactively altered to adjust the querying process in real-time.
Furthermore, the user can use the intermediate results to examine the querying process and comprehend the final answers.

\section{Related Work}
\label{section_related_work}

This section briefly introduces QA-related natural language interfaces (NLIs) and presents related works on QA over DBLP.

\textbf{NLIs for QA.}
NeuralQA~\cite{DBLP:conf/emnlp/Dibia20} is a library that provides a visual interface for entering questions and presenting gradient-based answer explanations.
While it leverages BERT~\cite{DBLP:conf/naacl/DevlinCLT19} to support a range of helpful QA functions (e.g., contextual query expansion), it does not allow for manual selection in entity linking, which is supported in NLQxform-UI.
IQA~\cite{DBLP:journals/ws/ZafarDLD20} is an existing system that allows the user to resolve entity linking by answering yes or no for generated candidates.
However, its applicability to DBLP is unclear.
With different target users, UKP-SQUARE~\cite{DBLP:conf/acl/BaumgartnerWSGE22} aims to provide researchers with an online platform to choose different available QA pipelines in a unified interface, enabling model exploration and comparison.

\textbf{QA on DBLP.}
To the best of our knowledge, NLQxform-UI is the first NLI over DBLP focusing on complex QA.
However, there are several QA methods that can be potentially extended for this purpose.
Jiang et al.~\cite{DBLP:conf/semweb/00010U23} propose a prompt-based approach that uses a pre-trained large language model to generate the structure and content of the SPARQL query for a given question.
BERTologyNavigator~\cite{DBLP:conf/semweb/RajpalU23} answers a question by navigating over the DBLP KG to extract correct relations and, consequently, answers.
PSYCHIC~\cite{DBLP:conf/semweb/Akl23} uses a neuro-symbolic approach to create an extractive QA model that can generate the SPARQL query for a given question.

\begin{figure*}[t]
    \centering
    \includegraphics[width=0.94\linewidth]{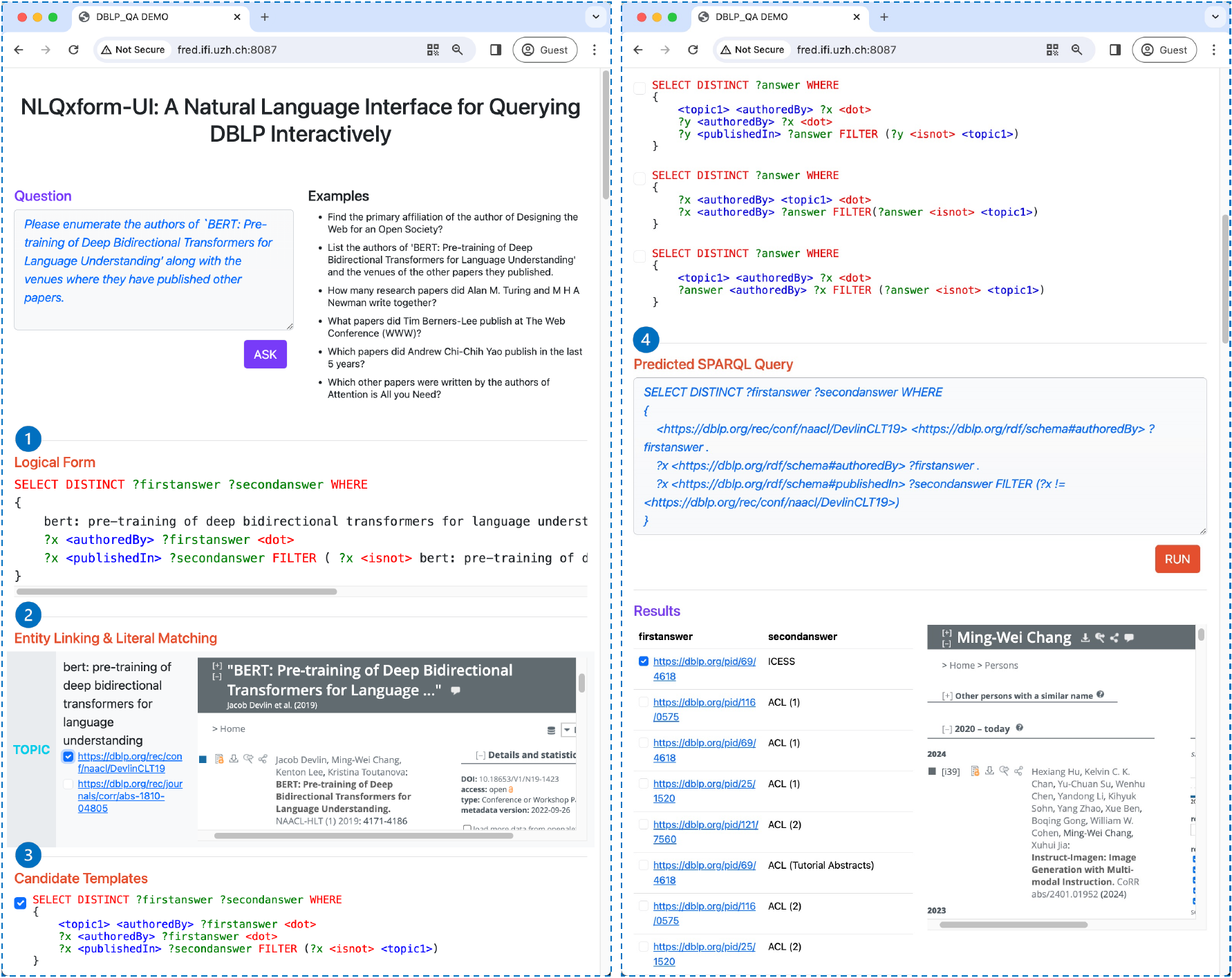}
    \caption{The user interface of NLQxform-UI, demonstrating the four-step querying process for the question ``\textit{please enumerate the authors of `BERT: Pre-training of Deep Bidirectional Transformers for Language Understanding' along with the venues where they have published other papers.}''}
    \label{figure_system}
\end{figure*}

\section{The NLQxform-UI System}
\label{section_the_nlqxform-ui_system}

In this section, we introduce NLQxform-UI with a specific use case---answering the example question:

\noindent``\textit{please enumerate the authors of `BERT: Pre-training of Deep Bidirectional Transformers for Language Understanding' along with the venues where they have published other papers.}''

The user interface of NLQxform-UI is presented in \cref{figure_system}, where the web page is split in the middle to accommodate the paper layout.
The field for question input (i.e., \textcolor{RoyalPurple}{\textbf{Question}}) is at the top of the web page.
Next to the input field, we also offer a few example questions (i.e., \textbf{Examples}) that the user can directly click on to use.
Given the above question, after clicking on the \textcolor{RoyalPurple}{\textbf{Ask}} button, the page will be redirected to the \textcolor{RoyalPurple}{\textbf{Results}} field, where the final answers are presented.
The first column (i.e., \textbf{first answer}) enumerates the URLs of the requested authors in DBLP (e.g., the URL of the second author of the BERT paper~\cite{DBLP:conf/naacl/DevlinCLT19}---\url{https://dblp.org/pid/69/4618}).
The second column (i.e., \textbf{second answer}) lists the corresponding venues where the authors have published other papers (e.g., ``ICESS'' and ``ACL'').
Next to the results, an inline frame is provided for the user to preview the answer URLs.
By default, the first URL is automatically previewed.
However, the preview would update in real-time when the user ticks a different URL.
Also, the user can freely scroll and open external links in the previewed page to explore other pages through the inline frame.
Compared to the static textual results of NLQxform and other QA methods without user interfaces (cf. \cref{section_related_work}), NLQxform-UI greatly facilitates viewing and examining the final answers.

Then, the user can scroll up to check other intermediate results of the querying process to better examine the validity of the returned answers.
Under \Circled{\textbf{1}} \textcolor{RedOrange}{\textbf{Logical Form}}, the result of the first step is presented:\footnote{Due to the size limitation, it is only partly displayed in \cref{figure_system}. This would not be an issue when using a normal-size browser window. Also, the user can use the scroll bar to read the complete content.}
\begin{Verbatim}[commandchars=\\\{\}]
\textcolor{Red}{SELECT} \textcolor{Red}{DISTINCT} \textcolor{OliveGreen}{?firstanswer} \textcolor{OliveGreen}{?secondanswer} \textcolor{Red}{WHERE}\\
\{\\
    the_BERT_paper \textcolor{blue}{<authoredBy>} \textcolor{OliveGreen}{?firstanswer} \textcolor{Red}{<dot>}\\
    \textcolor{OliveGreen}{?x} \textcolor{blue}{<authoredBy>} \textcolor{OliveGreen}{?firstanswer} \textcolor{Red}{<dot>}\\
    \textcolor{OliveGreen}{?x} \textcolor{blue}{<publishedIn>} \textcolor{OliveGreen}{?secondanswer} \textcolor{Red}{FILTER} 
    ( \textcolor{OliveGreen}{?x} \textcolor{Red}{<isnot>} the_BERT_paper )\\
\}
\end{Verbatim}
\noindent Please note that we use \verb|the_BERT_paper| to denote the long title of the BERT paper shown in \cref{figure_system}. The logical form has the same structure as the target SPARQL query and can be used to check if the query intention is precisely interpreted by the system.
It is generated by a BART~\cite{DBLP:conf/acl/LewisLGGMLSZ20} model in the text-to-text generation setup---translating given questions into logical forms.
We fine-tuned the BART using the DBLP-QuAD~\cite{DBLP:conf/birws/BanerjeeAUB23} dataset, which consists of 10,000 pairs of questions and SPARQL queries over the DBLP KG.
The basic elements of the SPARQL syntax (e.g., “SELECT”, “DISTINCT,” and “WHERE”), the frequently-used characters (e.g., parentheses and the dot), and the relation set of DBLP KG are added to the tokenizer of the BART.
Therefore, we can see that the above logical form already encompasses the SPARQL syntax and relation linking results.
However, the entities are still represented by their mentions in the given question.

Next, the system employs DBLP Search APIs\footnote{\url{https://dblp.org/faq/How+to+use+the+dblp+search+API.html}} to link the entity mentions in the logical form to DBLP URLs.
The results are presented under \Circled{\textbf{2}} \textcolor{RedOrange}{\textbf{Entity Linking \& Literal Matching}}.
Two candidate URLs are provided for the BERT paper.
The first links to the formal publication, while the second links to the pre-print version.
The user can check (or tick) and preview the URLs in the inline frame next to the links and make corrections in case of wrong initial selections.
Also, when choosing a different URL, the system automatically updates the query generation of the following steps in real time.

As the logical form is the direct output of the BART-based text generation, it may contain minor syntax errors (e.g., a missing or redundant parenthesis). 
The third step aims to utilize a pre-constructed template base~\cite{DBLP:conf/semweb/0003ZRRB23} to fix the errors.
We generated these templates by removing specific entities from the logical forms of the training questions in DBLP-QuAD.
For the above example question, NLQxform-UI retrieves a few candidate templates that are similar to the above logical form and presents them under \Circled{\textbf{3}} \textcolor{RedOrange}{\textbf{Candidate Templates}}. 
From \cref{figure_system}, we can see that the first one matches the logical form perfectly (\verb|the_BERT_paper| is represented by the placeholder \textcolor{blue}{<topic1>} in the template). 
Subsequently, NLQxform-UI utilizes the entity linking results to initialize the template, transforming it into an executable query shown under \Circled{\textbf{4}} \textcolor{RedOrange}{\textbf{Predicted SPARQL Query}}.
Furthermore, the user can manually edit the generated query in the input field and click on \textcolor{RedOrange}{\textbf{Run}} to execute it over DBLP KG through the query endpoint.\footnote{\url{https://dblp-kg.ltdemos.informatik.uni-hamburg.de/sparql}}
It is worth mentioning that other mismatching templates could also be useful for the user.
For example, the last candidate template shown in \cref{figure_system} can be used to generate a query that retrieves other papers published by the authors of the BERT paper.
NLQxform-UI enables the user to choose different candidate templates by ticking them, upon which the generated query would be updated in real-time.

\section{System Effectiveness and Usefulness}
\label{section_performance_evaluation}

NLQxform-UI performs the same as NLQxform if no manual changes are made through the interface.
NLQxform participated in the latest Scholarly QALD Challenge at the 22nd International Semantic Web Conference (ISWC 2023).\footnote{\label{challenge_web}\url{https://kgqa.github.io/scholarly-QALD-challenge/2023/}}
It is the winner of the task ``DBLP-QuAD --- Knowledge Graph Question Answering over DBLP''\footnote{\url{https://codalab.lisn.upsaclay.fr/competitions/14264}} with a significant improvement (+28.2\%) over the runner-up, regarding the F1 score computed based on answer precision and recall~\cite{DBLP:conf/semweb/0003ZRRB23}.
As the challenge has been the main platform for benchmarking QA over DBLP, NLQxform-UI possesses state-of-the-art QA effectiveness.

Also, we have explicitly evaluated NLQxform-UI following the setup of the challenge:
We used DBLP KG and 7,000/1,000 training/validation questions (with query and answer annotations) from DBLP-QuAD~\cite{DBLP:conf/birws/BanerjeeAUB23} to develop NLQxform-UI, including fine-tuning the BART model for the first step and constructing the template base for the third step.
Then, using the automatic evaluation platform,\footnote{\url{https://codalab.lisn.upsaclay.fr/competitions/14264\#participate}} we evaluated NLQxform-UI over 500 test questions.\footnote{\url{https://github.com/debayan/scholarly-QALD-challenge/blob/main/2023/datasets/codalab/finalphase/dblp-kgqa/dblp.heldout.500.questionsonly.json}} 
Without any manual changes, NLQxform-UI achieved the F1 score of 0.84.
In comparison, the next two existing best-performing methods Jiang et al.~\cite{DBLP:conf/semweb/00010U23} and BERTologyNavigator~\cite{DBLP:conf/semweb/RajpalU23} only achieved 0.66 and 0.22.

Moreover, NLQxform-UI is not limited to QA.
In real-world applications, it can be used as a web-based tool for entity linking when only given keywords.
Also, expert users could use it as an auxiliary tool for writing complex SPARQL queries---they can first use it to generate a draft query, then manually revise the draft, and finally check the results interactively in the interface.
Conference participants will be able to use the QA and all the above features of NLQxform-UI.

\section{Conclusion and Future Work}
\label{section_conclusion_and_future_work}

In this paper, we present a web-based natural language interface over DBLP: NLQxform-UI, which enables users to query DBLP using complex natural language questions through an intuitive and interactive web interface.
NLQxform-UI is powered by NLQxform, thus possessing the state-of-the-art effectiveness of NLQxform.
However, unlike NLQxform, the querying process of NLQxform-UI is re-organized into a human-on-the-loop process, where the intermediate results are presented to the user and can be altered in real-time.
This substantially improves the transparency of the system and facilitates users to examine and understand the returned answers.
In future work, we plan to adapt the system to other commonly used open-domain knowledge graphs.
The primary effort for this would involve further fine-tuning the adopted BART with other open-domain QA datasets.

\begin{acks}
This work was partially funded by the Digital Society Initiative of the University of Zurich, the University Research Priority Program ``Dynamics of Healthy Aging'' at the University of Zurich, 
and the Swiss National Science Foundation through Projects \href{https://data.snf.ch/grants/grant/184994}{``CrowdAlytics''} (Grant Number 184994) 
and \href{https://data.snf.ch/grants/grant/202125}{``MediaGraph''} (Grant Number 202125). 
\end{acks}

\bibliographystyle{ACM-Reference-Format}
\bibliography{bibfile}


\begin{thebibliography}{10}


\ifx \showCODEN    \undefined \def \showCODEN     #1{\unskip}     \fi
\ifx \showDOI      \undefined \def \showDOI       #1{#1}\fi
\ifx \showISBNx    \undefined \def \showISBNx     #1{\unskip}     \fi
\ifx \showISBNxiii \undefined \def \showISBNxiii  #1{\unskip}     \fi
\ifx \showISSN     \undefined \def \showISSN      #1{\unskip}     \fi
\ifx \showLCCN     \undefined \def \showLCCN      #1{\unskip}     \fi
\ifx \shownote     \undefined \def \shownote      #1{#1}          \fi
\ifx \showarticletitle \undefined \def \showarticletitle #1{#1}   \fi
\ifx \showURL      \undefined \def \showURL       {\relax}        \fi
\providecommand\bibfield[2]{#2}
\providecommand\bibinfo[2]{#2}
\providecommand\natexlab[1]{#1}
\providecommand\showeprint[2][]{arXiv:#2}

\bibitem[Akl(2023)]%
        {DBLP:conf/semweb/Akl23}
\bibfield{author}{\bibinfo{person}{Hanna~Abi Akl}.} \bibinfo{year}{2023}\natexlab{}.
\newblock \showarticletitle{{PSYCHIC:} {A} Neuro-Symbolic Framework for Knowledge Graph Question-Answering Grounding}. In \bibinfo{booktitle}{\emph{Joint Proceedings of Scholarly {QALD} 2023 and SemREC 2023 co-located with {ISWC} 2023, November 6-10, 2023}}, Vol.~\bibinfo{volume}{3592}.
\newblock


\bibitem[Banerjee et~al\mbox{.}(2023)]%
        {DBLP:conf/birws/BanerjeeAUB23}
\bibfield{author}{\bibinfo{person}{Debayan Banerjee}, \bibinfo{person}{Sushil Awale}, \bibinfo{person}{Ricardo Usbeck}, {and} \bibinfo{person}{Chris Biemann}.} \bibinfo{year}{2023}\natexlab{}.
\newblock \showarticletitle{DBLP-QuAD: {A} Question Answering Dataset over the {DBLP} Scholarly Knowledge Graph}. In \bibinfo{booktitle}{\emph{Proceedings of the 13th International Workshop on Bibliometric-enhanced Information Retrieval co-located with ECIR 2023, April 2nd, 2023}}, Vol.~\bibinfo{volume}{3617}. \bibinfo{pages}{37--51}.
\newblock


\bibitem[Baumg{\"{a}}rtner et~al\mbox{.}(2022)]%
        {DBLP:conf/acl/BaumgartnerWSGE22}
\bibfield{author}{\bibinfo{person}{Tim Baumg{\"{a}}rtner}, \bibinfo{person}{Kexin Wang}, \bibinfo{person}{Rachneet Sachdeva}, \bibinfo{person}{Gregor Geigle}, \bibinfo{person}{Max Eichler}, \bibinfo{person}{Clifton Poth}, \bibinfo{person}{Hannah Sterz}, \bibinfo{person}{Haritz Puerto}, \bibinfo{person}{Leonardo F.~R. Ribeiro}, \bibinfo{person}{Jonas Pfeiffer}, \bibinfo{person}{Nils Reimers}, \bibinfo{person}{G{\"{o}}zde~G{\"{u}}l Sahin}, {and} \bibinfo{person}{Iryna Gurevych}.} \bibinfo{year}{2022}\natexlab{}.
\newblock \showarticletitle{{UKP-SQUARE:} An Online Platform for Question Answering Research}. In \bibinfo{booktitle}{\emph{Proceedings of the 60th Annual Meeting of the Association for Computational Linguistics, {ACL} 2022 - System Demonstrations, May 22-27, 2022}}. \bibinfo{pages}{9--22}.
\newblock


\bibitem[Devlin et~al\mbox{.}(2019)]%
        {DBLP:conf/naacl/DevlinCLT19}
\bibfield{author}{\bibinfo{person}{Jacob Devlin}, \bibinfo{person}{Ming{-}Wei Chang}, \bibinfo{person}{Kenton Lee}, {and} \bibinfo{person}{Kristina Toutanova}.} \bibinfo{year}{2019}\natexlab{}.
\newblock \showarticletitle{{BERT:} Pre-training of Deep Bidirectional Transformers for Language Understanding}. In \bibinfo{booktitle}{\emph{Proceedings of the 2019 Conference of the North American Chapter of the Association for Computational Linguistics: Human Language Technologies, {NAACL-HLT} 2019, June 2-7, 2019, Volume 1}}. \bibinfo{pages}{4171--4186}.
\newblock


\bibitem[Dibia(2020)]%
        {DBLP:conf/emnlp/Dibia20}
\bibfield{author}{\bibinfo{person}{Victor Dibia}.} \bibinfo{year}{2020}\natexlab{}.
\newblock \showarticletitle{NeuralQA: {A} Usable Library for Question Answering (Contextual Query Expansion + {BERT)} on Large Datasets}. In \bibinfo{booktitle}{\emph{Proceedings of the 2020 Conference on Empirical Methods in Natural Language Processing: System Demonstrations, {EMNLP} 2020 - Demos, November 16-20, 2020}}. \bibinfo{pages}{15--22}.
\newblock


\bibitem[Jiang et~al\mbox{.}({[n.\,d.]})]%
        {DBLP:conf/semweb/00010U23}
\bibfield{author}{\bibinfo{person}{Longquan Jiang}, \bibinfo{person}{Xi Yan}, {and} \bibinfo{person}{Ricardo Usbeck}.} \bibinfo{year}{[n.\,d.]}\natexlab{}.
\newblock \showarticletitle{A Structure and Content Prompt-based Method for Knowledge Graph Question Answering over Scholarly Data}. In \bibinfo{booktitle}{\emph{Joint Proceedings of Scholarly {QALD} 2023 and SemREC 2023 co-located with {ISWC} 2023, November 6-10, 2023}}, Vol.~\bibinfo{volume}{3592}.
\newblock


\bibitem[Lewis et~al\mbox{.}(2020)]%
        {DBLP:conf/acl/LewisLGGMLSZ20}
\bibfield{author}{\bibinfo{person}{Mike Lewis}, \bibinfo{person}{Yinhan Liu}, \bibinfo{person}{Naman Goyal}, \bibinfo{person}{Marjan Ghazvininejad}, \bibinfo{person}{Abdelrahman Mohamed}, \bibinfo{person}{Omer Levy}, \bibinfo{person}{Veselin Stoyanov}, {and} \bibinfo{person}{Luke Zettlemoyer}.} \bibinfo{year}{2020}\natexlab{}.
\newblock \showarticletitle{{BART:} Denoising Sequence-to-Sequence Pre-training for Natural Language Generation, Translation, and Comprehension}. In \bibinfo{booktitle}{\emph{Proceedings of the 58th Annual Meeting of the Association for Computational Linguistics, {ACL} 2020, July 5-10, 2020}}. \bibinfo{pages}{7871--7880}.
\newblock


\bibitem[Rajpal and Usbeck({[n.\,d.]})]%
        {DBLP:conf/semweb/RajpalU23}
\bibfield{author}{\bibinfo{person}{Shreya Rajpal} {and} \bibinfo{person}{Ricardo Usbeck}.} \bibinfo{year}{[n.\,d.]}\natexlab{}.
\newblock \showarticletitle{BERTologyNavigator: Advanced Question Answering with BERT-based Semantics}. In \bibinfo{booktitle}{\emph{Joint Proceedings of Scholarly {QALD} 2023 and SemREC 2023 co-located with 22nd International Semantic Web Conference {ISWC} 2023, November 6-10, 2023}}, Vol.~\bibinfo{volume}{3592}.
\newblock


\bibitem[Wang et~al\mbox{.}({[n.\,d.]})]%
        {DBLP:conf/semweb/0003ZRRB23}
\bibfield{author}{\bibinfo{person}{Ruijie Wang}, \bibinfo{person}{Zhiruo Zhang}, \bibinfo{person}{Luca Rossetto}, \bibinfo{person}{Florian Ruosch}, {and} \bibinfo{person}{Abraham Bernstein}.} \bibinfo{year}{[n.\,d.]}\natexlab{}.
\newblock \showarticletitle{NLQxform: {A} Language Model-based Question to {SPARQL} Transformer}. In \bibinfo{booktitle}{\emph{Joint Proceedings of Scholarly {QALD} 2023 and SemREC 2023 co-located with 22nd International Semantic Web Conference {ISWC} 2023, November 6-10, 2023}}, Vol.~\bibinfo{volume}{3592}.
\newblock


\bibitem[Zafar et~al\mbox{.}(2020)]%
        {DBLP:journals/ws/ZafarDLD20}
\bibfield{author}{\bibinfo{person}{Hamid Zafar}, \bibinfo{person}{Mohnish Dubey}, \bibinfo{person}{Jens Lehmann}, {and} \bibinfo{person}{Elena Demidova}.} \bibinfo{year}{2020}\natexlab{}.
\newblock \showarticletitle{{IQA:} Interactive query construction in semantic question answering systems}.
\newblock \bibinfo{journal}{\emph{J. Web Semant.}}  \bibinfo{volume}{64} (\bibinfo{year}{2020}), \bibinfo{pages}{100586}.
\newblock


\end{thebibliography}

\end{document}